\documentclass[preprint]{aastex}

\shorttitle{FAST CMES WITHOUT X-RAY FLARES}

\shortauthors{Song et al.}

\begin{document}
\title{A Study of Fast Flareless Coronal Mass Ejections}
\author{H. Q. SONG\altaffilmark{1,3}, Y. CHEN\altaffilmark{1}, D. D. Ye\altaffilmark{1}, G. Q. HAN\altaffilmark{1}, G. H. Du\altaffilmark{1}, G. LI\altaffilmark{2,1}, J. ZHANG\altaffilmark{3}, AND Q. HU\altaffilmark{2}}

\affil{1 Institute of Space Sciences and School of Space Science
and Physics, Shandong University, Weihai, Shandong 264209, China}
\email{yaochen@sdu.edu.cn}

\affil{2 Department of Physics and CSPAR, University of Alabama in
Huntsville, Huntsville, AL 35899, USA}

\affil{3 School of Physics, Astronomy and Computational Sciences,
George Mason University, Fairfax, Virginia 22030, USA}

\begin{abstract}
Two major processes have been proposed to convert the coronal
magnetic energy into the kinetic energy of a coronal mass ejection
(CME): resistive magnetic reconnection and ideal macroscopic
magnetohydrodynamic instability of magnetic flux rope. However, it
remains elusive whether both processes play a comparable role or
one of them prevails during a particular eruption. To shed light
on this issue, we carefully studied energetic but flareless CMEs,
\textit{i.e.}, fast CMEs not accompanied by any flares. Through
searching the Coordinated Data Analysis Workshops (CDAW) database
of CMEs observed in Solar Cycle 23, we found 13 such events with
speeds larger than 1000 km s$^{-1}$. Other common observational
features of these events are: (1) none of them originated in
active regions; they were associated with eruptions of
well-developed long filaments in quiet-Sun regions, (2) no
apparent enhancement of flare emissions was present in soft X-ray,
EUV and microwave data. Further studies of two events reveal that
(1) the reconnection electric fields, as inferred from the product
of the separation speed of post-eruption ribbons and the
photospheric magnetic field measurement, were in general weak; (2)
the period with a measurable reconnection electric field is
considerably shorter than the total filament-CME acceleration
time. These observations indicate that, for these fast CMEs, the
magnetic energy was released mainly via the ideal flux rope
instability through the work done by the large scale Lorentz force
acting on the rope currents rather than via magnetic
reconnections. We also suggest that reconnections play a less
important role in accelerating CMEs in quiet Sun regions of weak
magnetic field than those in active regions of strong magnetic
field.
\end{abstract}


\section{Introduction}
Coronal mass ejections (CMEs) are the most energetic eruptions in
the solar system. Previous studies have established that many CMEs
are accompanied by flares and some of the physical properties of
CMEs and flares are closely related. For example, early
investigations using the Solar Maximum Mission (SMM) data showed
that about 48{\%} of CMEs were accompanied by X-ray flares
(Harrison 1995). The actual flare-accompanying rate could be
higher than 90{\%} if considering the fact that nearly half of the
observed CMEs were from the back side of the Sun. Later studies
revealed good correlations between CME speed (acceleration)
profiles with the soft X-ray (hard X-ray and microwave) profiles
of associated flares (e.g., Zhang et al. 2001; Qiu et al. 2004;
Mari\v ci\' c et al. 2007). In addition, the CME acceleration was
also found to be related to the inferred reconnection electric
fields (Qiu et al. 2004). Studies of connecting the extrapolated
magnetic flux in the flaring region to the magnetic flux of the
flux rope reconstructed from in-situ data showed that the two
fluxes were comparable (Qiu et al. 2007). These CME-flare
association studies strongly suggest an important role played by
magnetic reconnections in energizing CMEs.

Nevertheless, there are also studies indicating that in some
events CMEs and flares are only loosely connected. For instance,
it is well known that there exist CMEs without accompanying flares
or only associated with hardly-recognizable flares (e.g., Gosling
et al., 1976; Sheeley et al. 1999). On the other hand, $\sim$
70{\%} of C-level, 44{\%} of M-level, and 10{\%} of X-level flares
are not associated with CMEs according to recent statistical
studies (Yashiro et al. 2005; Wang and Zhang 2007). In addition,
statistical studies with a large sample of events showed that the
projected speed and kinetic energy of CMEs were only weakly
correlated with the peak values of soft X-ray flux (e.g., Yashiro
et al. 2002; Vr\v snak et al. 2005; see also Hundhausen (1997) for
an earlier study). After correcting the projection effect, the
correlation became even weaker (Yeh et al. 2005). An explanation
to these observations is that there exist other important factors
affecting the energy release process of CMEs besides magnetic
reconnection (e.g., Chen 2011). It should be noted that in many
studies magnetic reconnections were suggested to be the essential
energizing process of CMEs. This suggestion was often based on the
CME-flare synchronization and the evolutionary similarity of
specific physical quantities of the two phenomena, rather than on
the ground of direct cause-effect analysis.

Theoretically, it is well known that in the corona there exist two
important magnetic energy release mechanisms, one is the magnetic
reconnection, and the other the global magnetohydrodynamic (MHD)
flux rope instability (e.g., Van Tend and Kuperus 1978; Priest and
Forbes 1990; Forbes and Isenberg 1991; Isenberg et al. 1993;
Forbes and Priest 1995; Hu et al. 2003; Kliem and T\" or\" ok
2006; Fan and Gibson 2007; Chen et al. 2007a, 2007b; Olmedo and
Zhang, 2010). Via the later process the magnetic energy is
converted into the kinetic energy mainly through the work done by
the Lorentz force acting on the flux rope currents (c.f., Chen and
Krall, 2003; Chen et al. 2006, 2007a). Along with the rising of
the flux rope, a current sheet can be formed when field lines of
opposite directions are pushed together. This provides favorable
conditions for magnetic reconnection accounting for flares
(Carmichael 1964; Sturrock 1966; Hirayama 1974; Kopp \& Pneuman
1976). Magnetic energy can be further released through
reconnection to enhance the acceleration of the flux rope, giving
rise to a feedback relationship between the CME dynamics and
reconnection-induced processes (e.g., Zhang \& Dere 2006; Mari\v
ci\' c et al., 2007). Although this can explain the observed
CME-flare correlations, it does not promise the importance of
magnetic reconnection in CME energetics.

In a recent numerical simulation of CMEs, Chen et al. (2007a)
modeled the flux rope eruption within the framework of ideal MHD.
It was found that CMEs as fast as 1000 km s$^{-1}$ can be produced
even in the absence of magnetic reconnections. This point was
supported by a later numerical modeling of flux rope eruption
(Rachmeler et al. 2009). This naturally leads to a fundamental
question for CME studies: which energy release mechanism dominates
in a specific event, and what is the relative contribution of each
mechanism if both are important? The majority existing studies
focused on fast CMEs associated with flares from active regions.
In this study, we draw attentions to a special group of fast yet
flareless CMEs, trying to shed new light on the energizing process
of CMEs.

\section{Event identification and data analysis}

The events of this study were selected through the online CDAW
(Coordinated Data Analysis Workshops) database for CMEs observed
by the Large Angle and Spectrometric Coronagraph (LASCO; Brueckner
et al. 1995) on board the \textit{Solar and Heliospheric
Observatory (SOHO)} in Solar Cycle (SC) 23. We only considered
front-side flareless CMEs with a linear speed greater than 1000 km
s$^{-1}$ and an angular width larger than 20$^\circ$. The last
condition is to exclude the fast narrow jet events which are
considered to be physically different from normal CMEs (see, e.g.,
Chen, 2011). To ensure that the events are front-sided, we
examined the simultaneous images given by the Extreme Ultraviolet
Imaging Telescope (EIT, Delaboudiniere et al. 1995) on board
\textit{SOHO} to find clear eruptive signatures associated with
the CME, like post-eruption loops, ribbons, or eruptive filaments.
We require that these post-eruption signatures should be mostly
observed in the EIT field of view (FOV) to make sure that any
accompanying flare is not blocked by the solar disk. To determine
the presence of accompanying flares, we back-extrapolated the CME
start time assuming a constant propagation speed fixed at the
corresponding CME linear speed measured with the LASCO data, and
checked the GOES X-ray profile in a 6-hour time window centered at
the obtained start time. If a flare was found in this period, we
then verified its association with the CME using the EIT images
and the flare timing and source location provided by the National
Geophysical Data Center (NGDC)
(ftp://ftp.ngdc.noaa.gov/STP/space-weather/solar-data/). Only CMEs
without associated X-ray flares were included.

In total, we found 13 events conforming to the above conditions.
Through analyzing the EIT movies and available pre-eruption
H${\alpha}$ images, it was confirmed that all these events were
associated with eruptive filaments in the quiet-Sun region.
According to the magnetic field data given by the Michelson
Doppler Imager (MDI; Scherrer et al. 1995), we found that in 5
events the filaments were located above the neutral line inside a
single extended bipolar region (EBR), while in the rest 8 events
the filaments were located above the neutral region between two
adjacent EBRs. We refer the former 5 events as Group A, and the
rest 8 events as Group B. Similar classification method of CMEs
based on their source magnetic field distributions has been
adopted by Zhou et al. (2005, 2006).

In Table 1 we list the relevant parameters of the CMEs and
filaments. The event number and group identification are given in
the first column, the first appearance time of CME in the C2 FOV
and the back-extrapolated CME start time are given in the second
column. The third and fourth columns present the central position
angle (CPA) and the angular width of the CMEs, the fifth and sixth
columns show the projected linear speed and acceleration. The
estimated CME mass and kinetic energy are included in the seventh
and the eighth columns. All these data are taken from the CDAW
database. In the last columns we show the central position and
angular length of the pre-eruption filaments. These filament data
are provided by the Paris observatory
(http://bass2000.obspm.fr/home.php) except those for Events 1, 6,
and 13 whose parameters were measured by the authors.

\begin{table}[!htbp]
\begin{minipage}[t]{\columnwidth}
\caption{Parameters of the 13 CMEs and associated filaments, see
text for details.
 }
\renewcommand{\thefootnote}{\alph{footnote}}
\renewcommand{\footnoterule}{}
\tabcolsep=3pt
\begin{tabular}{lccccccrccc}
 \hline
 Event\#/ & \multicolumn{7}{c}{CME} & & \multicolumn{2}{c}{Filament} \\
 \cline{2-8}
 \cline{10-11}
Group & First C2 appe-  &  CPA & Ang.& Velocity & Acce. & Mass &
Ek & & Center & Ang.
\\ & arance time/CME & ($^{\circ}$) & Wid.($^{\circ}$) &  (km s$^{-1}$)  & (m s$^{-2}$)&
(g) & (erg) & & pos. ($^{\circ}$)& length ($^{\circ}$)
\\
 & start time (UT) & & & & & & & & &
\\
  \hline
01/B & 98/01/03 09:42/09:24 &  290  & 85  & 1020 & 21.8  &2.8e15 &1.4e31  & & N42W70 & 38 \\
02/B & 98/06/05 07:02/06:30 &  205  & 132 & 1017 & 28.3  &--     & --     & & S36W7  & 23 \\
03/B & 99/09/16 16:54/16:42 &  6    & 147 & 1021 & 49.3  &--     & --     & & N36W17 & 46 \\
04/B & 99/09/23 15:54/15:36 &  262  & 77  & 1150 & -5.8  &3.4e15 & 2.2e31 & & S23W41 & 22 \\
05/B & 00/07/24 16:54/16:29 &  220  & 50  & 1246 & -48.7 &3.2e15 & 2.4e31 & & S32W80 & 16\\
06/B & 02/03/02 15:06/15:00 &  124  & 149 & 1131 & 5.4   &--     & --     & & S28E74 & 48 \\
07/A & 02/05/11 20:06/19:47 &  19   & 52  & 1003 & 5.4   &1.1e15 & 5.5e30 & & N41E24 & 13 \\
08/A & 02/07/13 11:30/10:55 &  147  & 40  & 1037 & -21.6 &1.5e16 & 5.2e31 & & S31E28 & 18 \\
09/A & 02/08/06 18:25/17:35 &  218  & 134 & 1098 & -0.5  &--     & --     & & S36W25 & 46 \\
10/A & 02/12/21 02:30/02:04 &  2    & 225 & 1072 & -3.0  &--     & --     & & N40E9  & 24\\
11/B & 02/12/26 18:30/18:07 &  80   & 24  & 1075 & 5.7   &3.3e15 & 1.9e31 & & N14E73 & 7  \\
12/A & 03/01/05 10:37/09:52 &  353  & 67  & 1183 & 14.2  &1.3e15 & 9.2e30 & & N39E17 & 17\\
13/B & 05/01/04 09:30/09:14 &  288  & 102 & 1087 & 9.5   &6.9e15 & 4.1e31 & & N8W58  & 31 \\
  \hline
\end{tabular}
\end{minipage}
\end{table}

It can be seen from Table 1 that the average width of all CMEs is
$\sim$100$^\circ$, and the angular width is less than 50$^\circ$
in only two events, the CME linear speeds distribute in a range of
1003 -- 1246 km s$^{-1}$ with an average of 1088 km s$^{-1}$, the
acceleration varies between -48.7 and 49.3 m s$^{-2}$. Estimates
of the CME masses and kinetic energies are available from CDAW for
8 events. Among them, Event 8 (7) has the largest (smallest) mass
and energy being 1.5${\times}$10$^{16}$ (1.1${\times}$10$^{15}$g)
and 5.2${\times}$10$^{31}$ (5.5${\times}$10$^{30}$) erg. The
average values of the estimated CME masses and kinetic energies
for these 8 events are 4.63${\times}$10$^{15}$g and
2.33${\times}$10$^{31}$erg, $\sim$4 and 12 times larger than the
respective average values given by Webb et al. (2012) for all CMEs
observed from 1996 to 2011. It should be pointed out that large
errors are associated with these estimates of CME masses and
energies. Nevertheless, it is reasonable to say that most of our
events belong to relatively energetic ones among all CMEs. The
filament length varies in a range of 7${^\circ}$ - 48${^\circ}$
with an average of $\sim$27${^\circ}$. The filament CPAs
distribute in a range of latitudes from S36 -- N42.

In Figure 1 we present the radial variation of the projected
speeds of CMEs for Group A observed by LASCO C2 and C3 (left
panels), and the corresponding temporal evolution of the GOES
X-ray flux (0.1 - 0.8 nm) in the 6-hour time window centered at
the extrapolated CME start time (right panels). The corresponding
profiles for the 8 events in Group B are shown in Figure 2. The
CME speeds are given by the numerical differentiation using
3-point, Lagrangian interpolation. Note that the uncertainties of
the CME speed and acceleration come mainly from the uncertainty in
height measurements. The CME heights used here were measured by
the authors. Their measurement errors are estimated to be 5 pixels
in C2 and C3 images, corresponding to 0.0625 and 0.2938 R$_\odot$,
respectively. These errors are propagated in the standard way to
estimate the errors of velocity and acceleration. The obtained
speeds are not the same as that given by the CDAW database due to
the measurement uncertainty. In general our speeds are slightly
lower than those shown in CDAW. For example, the average of the
linear speeds in all events becomes 962 km s$^{-1}$, and the
speeds exceed 1000 km s$^{-1}$ in only 5 events. In the rest 5 and
3 events, the linear speeds are in the range of 900 - 1000 km
s$^{-1}$ and 800 - 900 km s$^{-1}$, respectively. Despite this
difference, all CMEs still belong to fast ones considering that
the average speed of CMEs observed from 1996 - 2006 with an
angular width larger than 30$^\circ$ was $\sim$470 km s$^{-1}$
(Gopalswamy et al. 2009). Therefore, it will not change the result
of this study.

In the 6-hour time window centered at the CME start time, there
are 23 GOES X-ray flares recorded by NGDC. These flares have been
listed in Table 2 of the Appendix, where we deduced that none of
them may have contributed significantly to the acceleration of our
CMEs, and thus are irrelevant to our study.

Comparing the left panels of Figures 1 and 2, we observe an
obvious difference of CME speed profiles. For 4 events in Group A
(except Event 10), the speeds keep almost constant or decline with
distance. This indicates that most events of this group have
finished the acceleration before entering into the C2 FOV (2
R$_\odot$ - 6 R$_\odot$). For 7 events in Group B (except Event
11), the speeds still increase significantly after entering into
the C2 FOV, and reach their peaks around 6 R$_\odot$ - 9
R$_\odot$. Similar speed differences of the two groups are also
present in the CDAW measurements. The group classifications stem
from the magnetic sources of the eruptions. We therefore suggest
that the observed speed evolutionary difference is related to
different magnetic sources on the photosphere and corresponding
large scale magnetic topologies in the corona. It should be
mentioned that the CME velocity measurements are accompanied by
uncertainties due to the projection effect, the instrument
sensitivity and cadence, and the subjective judgement of the
eruptive features. Therefore, the evolutionary speed difference
reported here should be verified in future studies.

We examined NGDC for possible microwave bursts in the obtained
6-hour time window of our events. The data are available for 9 of
our events occurring after Jan. 2000. It can be seen that none of
these events are associated with significant microwave
enhancements, say, reaching a level greater than 100 W
m$^{-2}\cdot$Hz$^{-1}$. Neither are seen considerable brightening
in the EUV wavelength according to the EIT data.

Now we summarize the common observational features of the 13
events as discussed above. All these CMEs are (1) front-side,
flareless, fast, and relatively-wide events, (2) accompanied by
post-eruption EUV loops, yet no considerable enhancements in
X-ray, EUV, and microwave data (when available), (3) originated
from quiet-Sun regions related to extended magnetic structures of
weak fields, (4) associated with well-developed filament
eruptions. Note the first group of features is given by our event
selection criteria. In the following discussion some of the above
features will be demonstrated with 2 events which are selected
from Group A (Event 10) and Group B (Event 2), respectively. In
both events the filament eruptions, and post-CME loops and ribbons
were clearly observed, microwave flux data at various frequencies
are also available from Nobeyama Solar Radio Observatory (NSRO).
In addition, both events are relatively close to the solar center
allowing the magnetic field strength to be measured with MDI.
Thus, besides demonstrating some of the above common observational
features of the 13 events, the two cases are also used to measure
the filament dynamics and estimate the reconnection electric
fields (e.g., Qiu et al. 2002).

The first appearance times of the two CMEs in C2 were 02:30 UT and
07:02 UT, and the back-extrapolated CME start times were 02:04 UT
and 06:30 UT. In Figure 3a we present the pre-eruption H$\alpha$
images of the filaments recorded by the Mauna Loa Solar
Observatory (MLSO), superposed by contours of the MDI magnetic
field data with yellow (blue) colors representing positive
(negative) magnetic polarities. The MDI synoptic map for CR1997 is
also shown to illustrate more clearly the relative location of the
filament (indicated by the red line) and the magnetic structure.
One C2 difference image and the EIT data of the post-eruption
loops are shown in Figures 3c and 3d. Similar data are shown in
Figure 4 for Event 2 of Group B. It can be seen that the filament
in Event 10 lies within an EBR region while that in Event 2
between two adjacent EBRs. Similar conclusions apply for the other
events in the respective group. In both events we can observe
diffusive signatures ahead of the bright CME ejecta and the
deflection of nearby coronal ray structures from Figures 3c and
4c. These features have been taken as possible signatures of
coronal shocks (e.g., Vourlidas et al. 2003; Sheeley et al. 2000;
Feng et al. 2013). Considering both CMEs are fast, it is
conceivable that coronal shocks were generated in both events.

It should be stressed that the post-eruption loops and ribbons
were observed in all our events. These flare-like features can be
taken as evidences of magnetic reconnection. In Figures 5 and 6 we
plot the NSRO microwave fluxes at frequencies of 2 (red), 3.75
(green) and 9.4 (blue) GHz in the obtained 6-hour time window for
Event 10 and Event 2, respectively. It can be seen that there were
no obvious responses to the eruptions in these microwave data, as
mentioned before. This indicates a lack of efficient plasma
heating and electron acceleration during the eruptions, although
reconnections are involved. This statement applies to all the
events with microwave data.

The Mark-IV (MK4) coronagraph (Elmore et al. 2003) observes the
corona with a smaller FOV than LASCO C2. However, none of our
events were well observed by MK4. Therefore, we made use of the
filament motion to approximate the CME dynamics in the lower
corona. Following Wang et al. (2003) and Qiu et al. (2004), we
measured the positions of the filament center and top, and deduced
the corresponding heights assuming a radially-outward propagation.
This reduces the projection effect on the height measurement. The
obtained filament heights are given in Figures 7 (Event 10) and 8
(Event 2) along with the heights of the CME front (measured by the
authors). The velocities and accelerations as well as their errors
were determined with the same method as used in Figures 1 and 2.
The filament height measurement uncertainty is estimated to be 5
pixels of the EIT image. This already considers the errors caused
by the projection effect. It can be seen that for Event 10 the
maximum velocities (accelerations) of the filament and CME fronts
are 140 km s$^{-1}$ (0.061 km s$^{-2}$) and 992 km s$^{-1}$ (0.073
km s$^{-2}$), respectively. For Event 2 the maximum velocities
(accelerations) are 136 km s$^{-1}$ (0.082 km s$^{-2}$) and 1083
km s$^{-1}$ (0.112 km s$^{-2}$) for the filament and CME front,
respectively. The vertical dashed lines in both figures represents
the time at which the post-eruption ribbons appeared in the FOV
and the time at which the separation stopped. These times are
02:00 (06:00) UT and 04:36 (08:05) UT for Event 10 (2). It can be
seen that the filament started to rise rapidly at $\sim$01:00
(05:00) UT, and the CME stopped the acceleration at $\sim$05:00
(09:00) UT in Event 10 (2). In other words, in both events the
filament rose earlier than the appearance of post-eruption
ribbons, and the total acceleration time of filaments and CME
fronts was considerably longer than the ribbon separation time.

Following Qiu et al. (2002, 2004) and Wang et al. (2003), we also
estimated the reconnection electric field, which is given by the
product of separation speeds of the EIT observed post-eruption
ribbons and the magnetic field strength measured with MDI. This
was done for three points along the ribbons (marked as A, B, and C
in Figures 3d and 4d). The average reconnection fields are
estimated to be 0.04, 0.03, 0.03 V cm$^{-1}$ for the three points
with a maximum less than 0.07 V cm$^{-1}$ in Event 10. These
values are 0.06, 0.04, 0.08 V cm$^{-1}$ with a maximum less than
0.1 V cm$^{-1}$ in Event 2. Nevertheless, due to the noise level
of the MDI magnetic field measurement, the error of the
reconnection electric field measurement is notoriously large
($\sim$ 0.1 V cm$^{-1}$) which is actually comparable to the
physical values themselves. This prevents us from getting any
meaningful conclusion of the variation trend of the electric
field. The only robust deduction is that the reconnection electric
field remains weak during the reconnection process. This is
consistent with the lack of significant emission enhancements in
the X-ray, EUV, and microwave wavelength. The inferred
reconnection electric field does not exceed 0.1-0.2 V cm$^{-1}$,
much weaker than the values reported in previous studies. For
example, Qiu et al. (2004) reported that the maximum reconnection
electric field, estimated with the same methods as that used here,
was 5.8 (0.5) V cm$^{-1}$ for a CME with an accompanying X1.6
(M1.0) flare.

\section{Discussion}
It is a core issue of CME studies regarding the release mechanism
of magnetic energy in the corona. Two major mechanisms have been
suggested including the ideal MHD flux rope instability and
magnetic reconnection (see, e.g., Chen et al., 2007a, 2007b; Chen
2013). It is possible that in a specific event both processes
contribute. Then a natural question arises: how can we determine
the relative contribution of each mechanism? Observationally this
is not a trivial issue to resolve due to the following
limitations. (1) There is no direct technique to measure the
coronal magnetic field on a daily basis, and therefore the energy
involved in the eruption can not be determined directly; (2) In
many events the dynamics of CMEs in the inner corona can not be
observed due to the FOV limits of coronagraphs, while a major part
of the CME acceleration (as well as the energy conversion) takes
place there; (3) Available estimates of the CME initial
acceleration, mass and kinetic energy suffer from large errors
because of the instrument sensitivity and resolution, the
projection effect, as well as the limitations of the plasma
density inversion method. These factors make the studies of CME
energetics difficult.

Theoretically, the relative contribution of the two processes can
be evaluated with a numerical model of flux rope eruption as done
by Chen et al. (2007a). As mentioned previously, Chen et al. found
that a flux rope eruption as fast as 1000 km s$^{-1}$ can be
simulated with the ideal flux rope instability in the absence of
magnetic reconnection. After reconnection sets in, the flux rope
moves faster with more magnetic energy being converted into the
CME kinetic energy. They compared the kinetic energy changes of
the system with and without reconnections and concluded that the
two magnetic energy release processes, if reconnections are
invoked, can have comparable contributions to the flux rope
eruption. Note that numerical resistivity was used to invoke
reconnections in their model, which is believed to be several
orders of magnitude larger than a physical one. Therefore, the
obtained contribution of reconnection to the increase of the flux
rope kinetic energy should be taken as an upper limit.

Generally speaking, the contribution of different mechanism in a
specific event shall depend on the details of the global and local
magnetic topologies, as well as the details of reconnections and
the eruption. For the 13 events listed in Table 1, although they
belong to fast CMEs with kinetic energies estimated to be
considerably larger than average values, no accompanying X-ray
flares were observed, neither the usual flare-related emission
fluxes in the soft-X ray and microwave wavelengths showed
appreciable enhancement. The reconnection electric fields
estimated for two events remained lower than 0.1 V cm$^{-1}$
during the eruption. In addition, in both events the filament rose
$\sim$1h earlier than the appearance of post-eruption ribbons, and
the total acceleration time of filaments and CMEs was considerably
longer than the time with a measurable reconnection electric
field. These observational features strongly indicate that
magnetic reconnections played only a minor role in accelerating
the CMEs out of the corona. Considering that (1) reconnections and
flux rope instabilities are the two major magnetic energy release
mechanisms proposed so far, and (2) all our events correspond to
well-developed filament eruptions indicating the presence of
large-scale flux rope structure (e.g., Low 1996), we suggest that
the flux rope instability dominates the energy releases in our
events.

The above judgement on the existence of flux ropes in the corona
is mainly based on the presence of filaments. Yet, it is not
possible to confirm the existence of such a coronal magnetic
structure observationally as the coronal magnetic field can not be
measured directly at this time. There exist other observational
signatures supporting the existence of a flux rope in the corona,
like the rope-like or tangled white light structures from the
coronagraph observations (e.g., Dere et al. 1999), the sigmoid
structure observed in the CME source (Canfield et al. 1999;
McKenzie {\&} Canfield 2008) as well as coronal cavities (e.g.,
Gibson et al. 2006; Riley et al. 2008). In recent studies using
the high-resolution \textit{SDO} (Solar Dynamics Observatory)
data, a hot-channel structure was observed in the 131 \AA\
wavelength (Cheng et al. 2011; Zhang et al. 2012) and interpreted
as a kind of flux rope structure. According to Zhang et al. (2012)
the hot channel manifested itself as a twisted structure which got
untwisted to an accelerating and expanding arcade. It is obvious
that the active-region hot-channel flux rope is different in
thermal and magnetic properties from those quiet-Sun filament-flux
rope structure discussed in the present study. It remains unknown
whether and how these discrepancies of flux rope structures may
affect the contributions of different energy release process.

It is useful to compare and understand why flareless fast CMEs are
very different from those fast events with strong accompanying
flares. The fast events with strong flares usually originate from
solar active regions, while all our events are from EBRs with
relatively weak fields in the quiet-Sun region. This difference of
magnetic sources may be one important cause of the different
characteristics of the above two types of fast CMEs. As suggested,
magnetic reconnections play only a minor role in our events, whose
role shall presumably become more important in those fast CMEs
with strong flares. Yet, it is not possible to determine the
relative contribution of reconnections and flux rope instabilities
with available observations. It is also not clear whether the
above two types of fast CMEs are physically different or they just
represent two extremes of a broad CME distribution with similar
eruption mechanism(s). This issue should be addressed in future
studies.

Another interesting point is that for all our events there were no
accompanying type II radio bursts according to the Radio Solar
Telescope Network (RSTN) and NGDC data, neither were there
associated solar energetic particle (SEP) events based on the CDAW
database. As mentioned, the CMEs propagated very fast with a high
possibility of driving a coronal shock. And we indeed observed in
$\sim$10 of our events the diffusive halo structure ahead of the
bright CME ejecta and the deflections of coronal streamers that
were not in apparent contact with the ejecta. These phenomena are
regarded as possible shock signatures (e.g., Vourlidas et al.
2003; Sheeley et al. 2000, Feng et al., 2013). In addition, among
our events 7 were from the western hemisphere. This favors the
propagation of energetic particles to the near-Earth space.
Therefore, one natural explanation for the absence of type IIs and
SEPs is that the shocks, if they existed, were not capable of
accelerating electrons and ions to high enough energies.

Relevant studies indicate that efficient particle acceleration via
shocks may require seed particles with enough number and energy,
as well as an enhanced turbulence level. These conditions may be
provided by the accompanying flare process or by preceding
eruptions (e.g., Li 2012; Ding et al., 2013). With a careful
examination, it was found that in 10 of the 13 events there seem
no preceding eruptions that were able to strongly disturb the
propagation environment of the CME in study. For the rest 3 events
(Event 2, 5, and 6), preceding-CMEs were observed. This possibly
indicates that preceding-CMEs are only one favorable condition for
efficient particle acceleration. In summary, the above analysis
implicates that the conditions of no accompanying flares and no
preceding-CMEs do not favor the acceleration of energetic
particles via shocks, thus provides a support to the above
theories of shock particle acceleration. Of course, it is not
possible to completely rule out the possibility that there were
energetic particles and type IIs associated with the shock, yet
not observable due to their propagation and transport effect.

\section{Summary}
In this paper we presented a special group of fast front-side
CMEs, which had no associated flares. We found that all the events
were associated with filament eruptions originated either within a
single EBR or between adjacent EBRs in the quiet-Sun region.
During the eruption, we observed post-eruption ribbons and loops,
in accordance with the standard flare picture, indicating the
presence of magnetic reconnections in these events. However the
following features suggest that ideal MHD instabilities of the
flux rope rather than reconnections played a dominant role in the
energy release process of the events. (1) No accompanying GOES
X-ray flares and corresponding flux enhancements in the soft X-ray
and microwave radiations were present. This means that the
relevant radiations of this event were below the background
emission level. (2) In two events, it was revealed that the
filaments rose earlier than the appearance of post-eruption
ribbons, and the total acceleration time of filaments and CME
fronts was considerably longer than the time with measurable
reconnection electric field that was weak in general. (3) The
presence of well-developed long filaments indicated the existence
of large-scale flux rope structures in the corona. These results
are consistent with previous modelling results showing that a fast
eruption of the flux rope can be yielded considering the ideal
flux rope instabilities as the only magnetic-energy release
mechanism of the eruption, and suggest that magnetic reconnection
plays a less important role in accelerating CMEs in quiet Sun
regions of weak magnetic field than those in active regions of
strong magnetic field.

\section{Appendix}
In this section, we present all the 23 flares reported by the
National Geophysical Data Center (NGDC)
(ftp://ftp.ngdc.noaa.goc/STP/spaceweather/solar-data/) in the
6-hour time window centered at the back-extrapolated CME start
time, and determine their relevance to the CME major
accelerations. These flares have been numbered in the GOES X-ray
fluxes (0.1 - 0.8 nm) shown in Figures 1 and 2. In table 2 we list
the relevant information of the flares. The flare and event
numbers and group identification are given in the first column.
The start, peak and stop time are given in the next three columns.
The last two columns are their X-ray classifications and
locations, including the active region number. There are 6 flares
whose source information remains unknown. Parameters listed here
are taken from NGDC.

\begin{table}[!htbp]
\begin{minipage}[t]{\columnwidth}
\caption{Information of the 23 flares numbered in Figures 1 and 2.
}

\renewcommand{\thefootnote}{\alph{footnote}}
\renewcommand{\footnoterule}{}
\tabcolsep=3pt
\begin{tabular}{ccccccc}
 \hline
 Flare\#/Event\# & Start Time  & Peak Time & Stop Time & Class &
 Position (NOAA)\\
 /Group & (UT) & (UT) & (UT) & & &
\\
 \hline
01/01/B & 98/01/03 12:12 & 12:20 & 12:26 & C3.3 & No Source
Info.\footnotemark[1]\footnotetext[1]{Flares started 1 hour
after the CME appearance time in LASCO C2.}\\
02/02/B & 98/06/05 04:39 & 04:45 & 04:51 & C1.4 & S26E43
(08232)\footnotemark[2]\footnotetext[2]{Flares ended
before the apparent rise of the associated filament.}\\
03/02/B & 98/06/05 06:43 & 06:48 & 06:54 & B9.0 & S25E45 (08232)\\
04/05/B & 00/07/24 15:55 & 16:00 & 16:21 & C2.8 & S31E45 (09100)\\
05/06/B & 02/03/02 12:55 & 12:59 & 13:03 & C1.1 & No Source Info.\footnotemark[2]\\
06/07/A & 02/05/11 16:45 & 16:50 & 16:54 & C2.7 & S06E48 (09946)\footnotemark[2]\\
07/07/A & 02/05/11 17:32 & 17:38 & 17:41 & C3.7 & S17W58 (09934)\footnotemark[2]\\
08/07/A & 02/05/11 18:32 & 18:36 & 18:37 & B9.7 & S15W57 (09934)\footnotemark[2]\\
09/07/A & 02/05/11 21:46 & 21:50 & 21:53 & C1.5 & No Source Info.\footnotemark[1]\\
10/08/A & 02/07/13 08:12 & 08:15 & 08:18 & C2.2 & N17E35 (10030)\footnotemark[2]\\
11/09/A & 02/08/06 15:16 & 15:27 & 15:52 & C7.6 & S06W66 (10057)\footnotemark[2]\\
12/09/A & 02/08/06 20:05 & 20:09 & 20:11 & C1.3 & No Source Info.\footnotemark[1]\\
13/10/A & 02/12/20 23:17 & 23:21 & 23:24 & C2.0 & S25W37 (10226)\footnotemark[2]\\
14/10/A & 02/12/20 23:58 & 00:02 & 00:08 & C1.7 & S28W38 (10226)\footnotemark[2]\\
15/10/A & 02/12/21 02:46 & 02:52 & 02:57 & C2.4 & S27W45 (10226)\\
16/10/A & 02/12/21 03:39 & 03:49 & 03:56 & C2.9 & N19W19 (10229)\footnotemark[1]\\
17/10/A & 02/12/21 04:38 & 04:41 & 04:45 & C2.2 & No Source Info.\footnotemark[1]\\
18/10/A & 02/12/21 04:48 & 04:53 & 04:57 & C4.6 & S25W41 (10226)\footnotemark[1]\\
19/12/A & 03/01/05 07:54 & 07:58 & 08:02 & C1.0 & S08E27 (10242)\footnotemark[2]\\
20/12/A & 03/01/05 09:24 & 09:28 & 09:32 & B9.2 & S19W55 (10243)\\
21/12/A & 03/01/05 11:51 & 11:54 & 11:58 & C1.3 & No Source Info.\footnotemark[1]\\
22/13/B & 05/01/04 05:59 & 06:16 & 06:30 & C3.3 & N05W07 (10715)\footnotemark[2]\\
23/13/B & 05/01/04 10:53 & 11:13 & 11:29 & C7.3 & N05W11 (10715)\footnotemark[1]\\
  \hline
\end{tabular}
\end{minipage}
\end{table}

The flare-CME associations are determined via the following three
steps. Firstly, flares started 1 hour after the CME appearance
time in LASCO C2 are considered to be irrelevant. At this time,
the CME, propagating very fast in general, should have moved to
the outer corona and finished most of its acceleration. This
removes 11 flares as relevant events, including 5 flares without
source information. They are indicated with a subscript ``a' in
the table. Further examinations of the LASCO data confirmed that
the CME fronts are already beyond 8 R$_\odot$ at the time of these
flares. Secondly, flares ended before the apparent rise of the
associated filament, marked with a subscript ``b', are also
considered to be irrelevant. This removes 8 other flares including
the last one with unknown source location. Last, we examined the
left 4 flares (No. 3, 4, 15, and 20), and found that their X-ray
emissions are only marginally beyond the background emission
level, and they are from active regions not close to our CME
sources as verified by the EIT imagings. Therefore, neither are
they considered to be relevant to our study.

\acknowledgments

We thank the anonymous referee for his/her constructive comments
that improved this paper.  We are grateful to Dr. Guangli Huang,
Jun Zhang, Guiping Zhou, and Xin Cheng for their comments and
suggestions. This work was supported by the 973 program
2012CB825601, NNSFC grants 41104113, 41274177, and 41274175. H. Q.
Song was also supported by the Natural Science Foundation of
Shandong Province ZR2010DQ016, and the Independent Innovation
Foundation of Shandong University 2010ZRYB001. We acknowledge the
use of the CDAW CME catalog, which is generated and maintained at
the CDAW Data Center by NASA and the Catholic University of
America in cooperation with the Naval Research Laboratory. SOHO is
a project of international cooperation between ESA and NASA. We
are also grateful to the NSRO, MLSO, BBSO, KSO, and YNAO teams for
making their data available to us.

\clearpage

\begin{figure}
\epsscale{1.0} \plotone{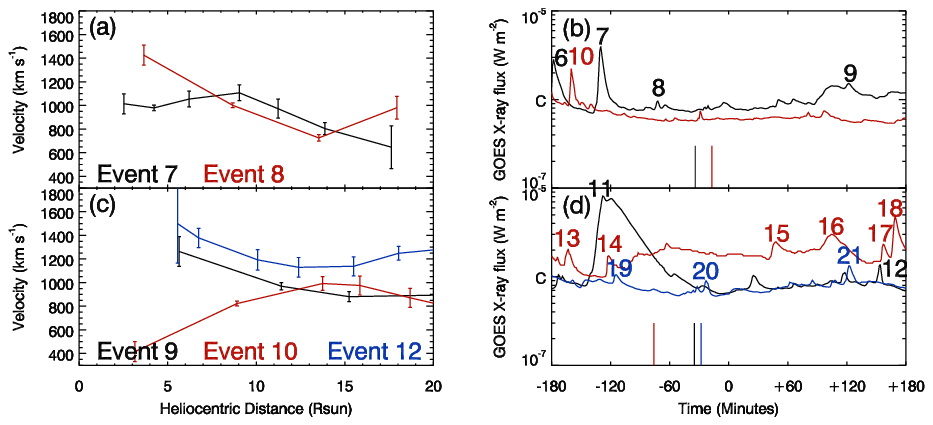} \caption{CME speeds and GOES
X-ray fluxes for the 5 Group A events listed in Table 1. Left
panels are the radial variation of the projected speeds of CMEs
measured with the LASCO C2 and C3 data. Right panels are the
corresponding GOES X-ray (0.1 - 0.8 nm) fluxes in the 6-hour time
window centered at the back-extrapolated CME start time. Flares
given by NGDC in the time window are numbered, whose information
is listed in Table 2 of the Appendix. Colors are used to represent
different CME events. The vertical lines in the right panels
indicate the time the associated filaments start to rise
noticeably. (A color version of this figure is available in the
online journal.)\label{fig1}}
\end{figure}

\begin{figure}
\epsscale{1.0} \plotone{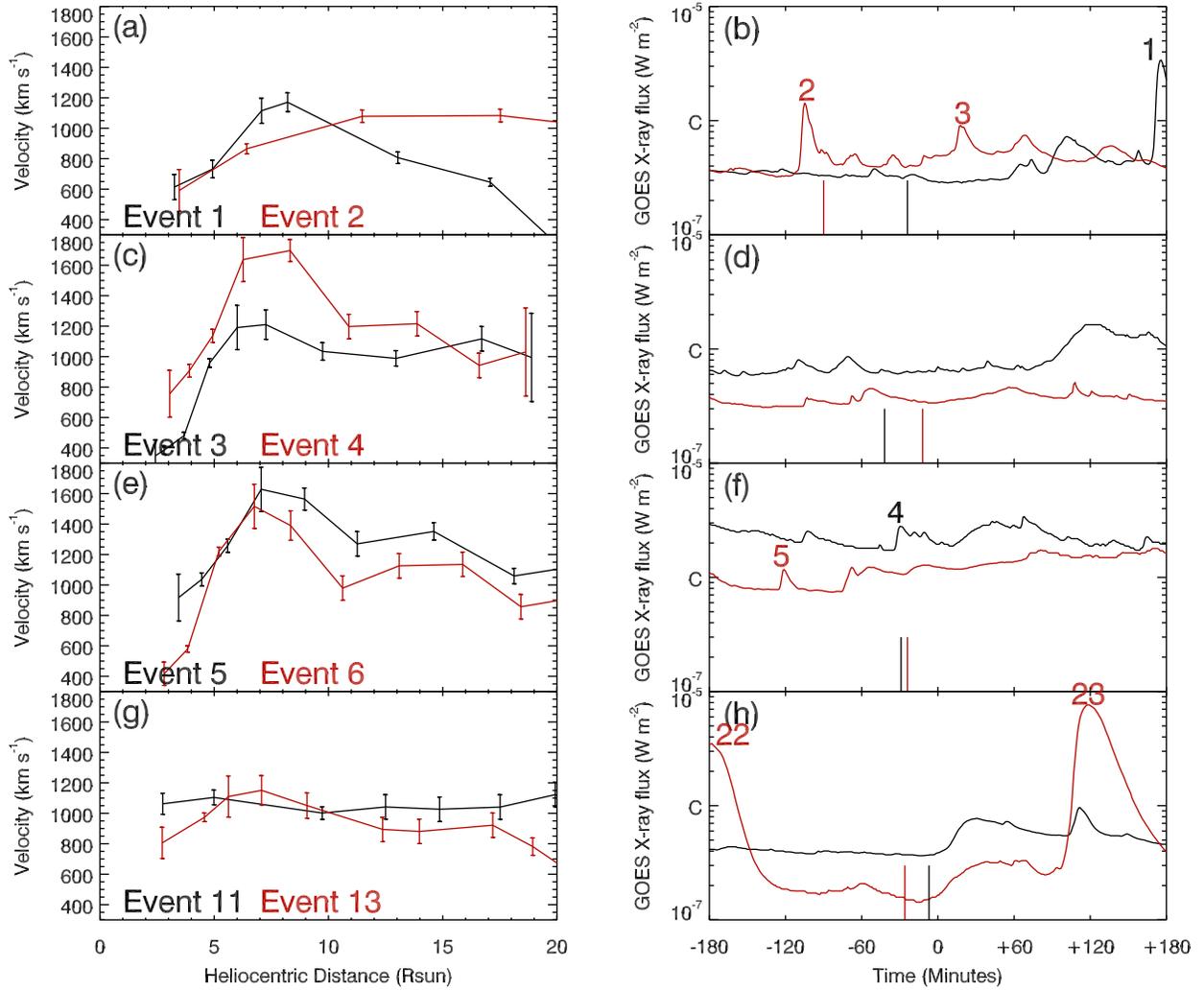} \caption{Same as Figure 1 but
for the 8 Group B events listed in Table 1. (A color version of
this figure is available in the online journal.)\label{fig2}}
\end{figure}

\begin{figure}
\epsscale{.80} \plotone{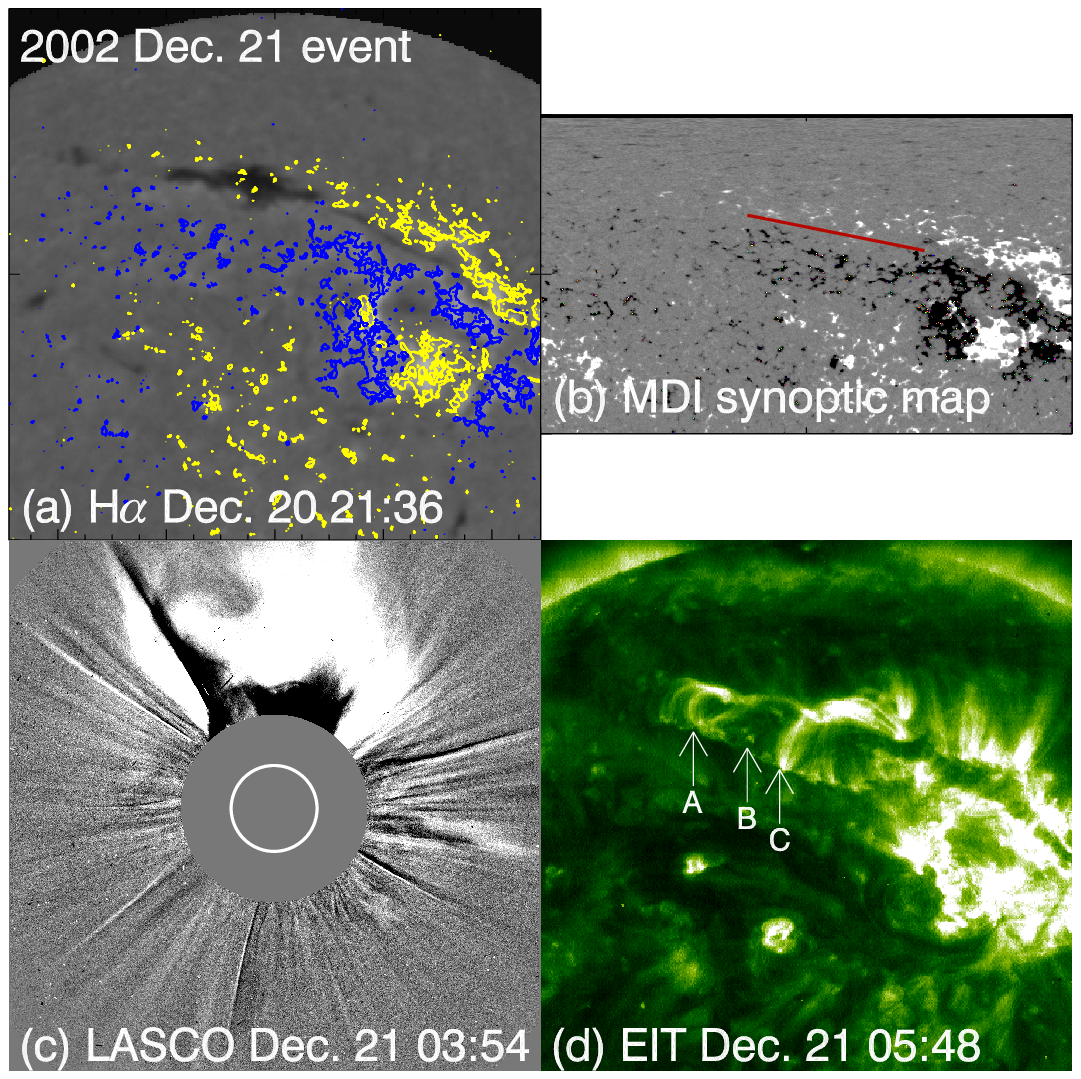} \caption{Observations of the
2002 Dec. 21 event (Event 10) in Group A. (a) The pre-eruption
H$\alpha$ images recorded by the MLSO, superposed by contours of
the MDI magnetic field data (20:48 UT on the same day) with yellow
(blue) colors representing positive (negative) magnetic
polarities. (b) MDI synoptic map for CR1997 with a FOV of [0, 90]
for the latitude and [30, 120] for the longitude, the red line
indicates the location of the relevant filament. (c) The LASCO C2
difference image of the CME. (d) the EIT data of the post-eruption
loops. The FOVs of (a) and (d) are identical taken to be [-0.5
R$_\odot$, 0.5 R$_\odot$] for the horizontal axis and [0, 1
R$_\odot$] for the vertical axis. Three points in (d) along the
expanding loops are marked for reconnection election field
measurements. (Animations of the EIT data and a color version of
this figure is available in the online journal.)\label{fig3}}
\end{figure}

\begin{figure}
\epsscale{.80} \plotone{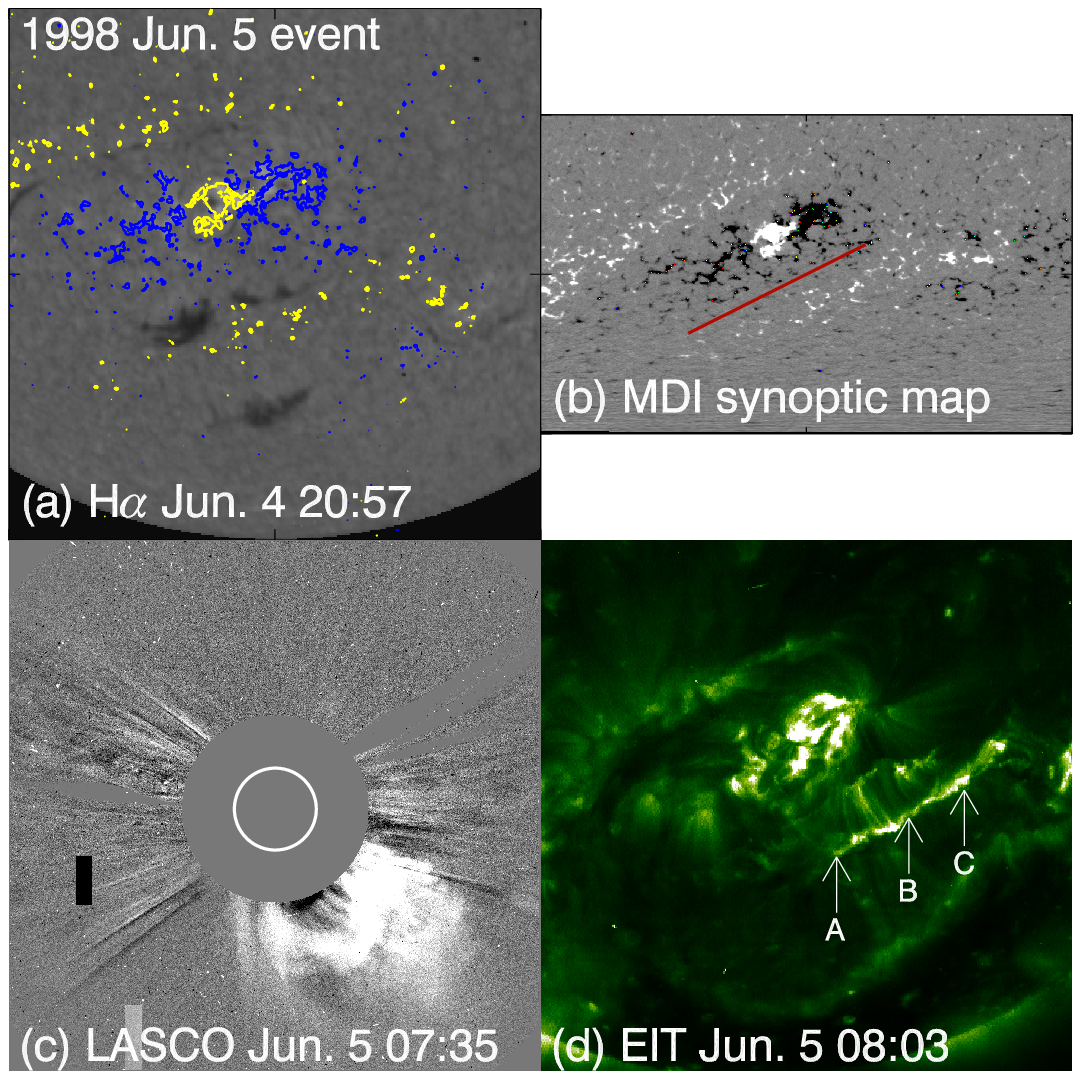} \caption{Observations of the
1998 Jun. 5 event (Event 2) in Group A. (a) The pre-eruption
H$\alpha$ images recorded by the MLSO, superposed by contours of
the MDI magnetic field data (20:48 UT on the same day) with yellow
(blue) colors representing positive (negative) magnetic
polarities. (b) MDI synoptic map for CR1936 with a FOV of [-90, 0]
for the latitude and [0, 90] for the longitude, the red line
indicates the location of the relevant filament. (c) The LASCO C2
difference image of the CME. (d) the EIT data of the post-eruption
loops. The FOVs of (a) and (d) are identical taken to be [-0.5
R$_\odot$, 0.5 R$_\odot$] for the horizontal axis and [-1, 0
R$_\odot$] for the vertical axis. Three points in (d) along the
expanding loops are marked for reconnection election field
measurements. (Animations of the EIT data and a color version of
this figure is available in the online journal.)\label{fig4}}
\end{figure}

\begin{figure}
\epsscale{.80} \plotone{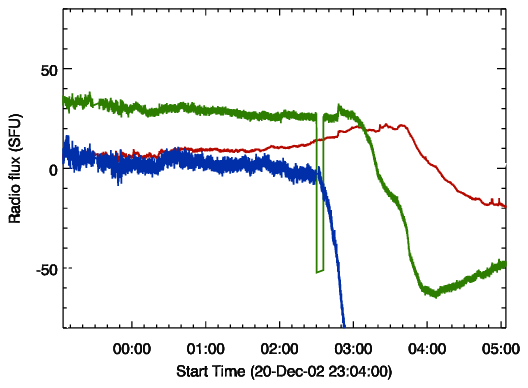} \caption{The NSRO microwave
fluxes at frequencies of 2 (red), 3.75 (green) and 9.4 (blue) GHz.
The center of the 6-hour time window is the estimated CME start
time. (A color version of this figure is available in the online
journal.)\label{fig5}}
\end{figure}

\begin{figure}
\epsscale{.80} \plotone{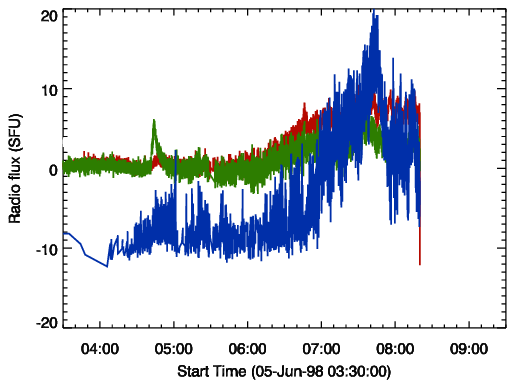} \caption{The NSRO microwave
fluxes at frequencies of 2 (red), 3.75 (green) and 9.4 (blue) GHz.
The center of the 6-hour time window is the estimated CME start
time. (A color version of this figure is available in the online
journal.)\label{fig2}}
\end{figure}

\begin{figure}
\epsscale{0.80}
 \plotone{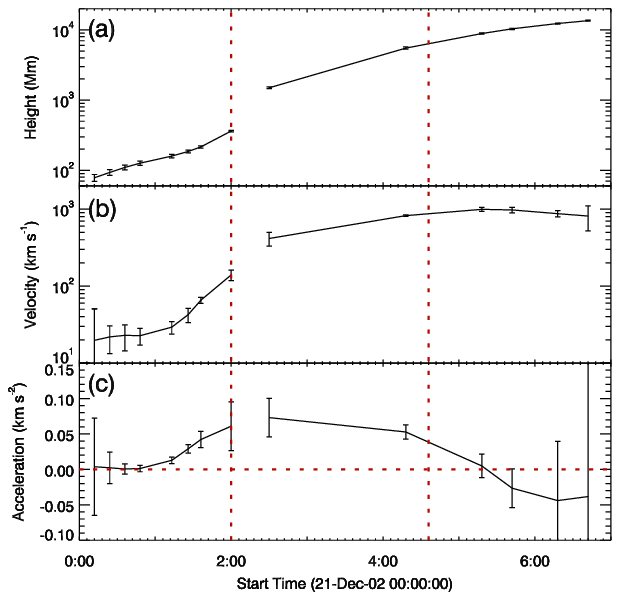}
\caption{Heights, velocities, and accelerations of the filaments
and CME fronts for the 2002 Dec. 21 event (Event 10) in Group A.
See text for details. The vertical dashed lines indicate the time
at which the post-eruption ribbons appear and the time at which
they stop separating from each other. The horizontal line depicts
the zero acceleration. \label{fig7} (A color version of this
figure are available in the online journal.)}\end{figure}

\begin{figure}
\epsscale{0.80}
 \plotone{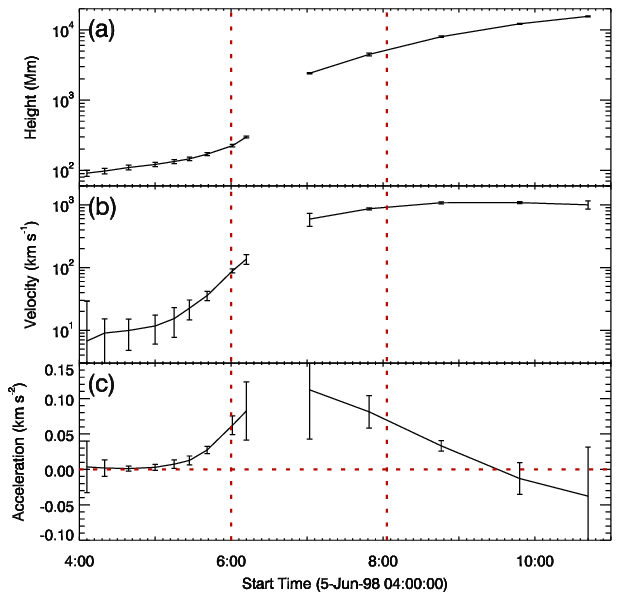}
\caption{Heights, velocities, and accelerations of the filaments
and CME fronts for the 2002 Dec. 21 event (Event 10) in Group A.
See text for details. The vertical dashed lines indicate the time
at which the post-eruption ribbons appear and the time at which
they stop separating from each other. The horizontal line depicts
the zero acceleration.\label{fig8} (A color version of this figure
are available in the online journal.)}
\end{figure}
\end{document}